\documentclass[a4paper,10pt]{article}
\usepackage[utf8x]{inputenc}
\usepackage{graphicx}
\usepackage{amsmath}
\usepackage{cite}
\usepackage{authblk}
\usepackage[margin=3cm]{geometry}

\title{\bf Very early warning signal for El~Ni\~no in 2020 with a 4 in 5 likelihood}
\date{}

\author[1] {Josef Ludescher}
\author[2] {Armin Bunde}
\author[3] {Shlomo Havlin}
\author[1] {Hans Joachim Schellnhuber}

\tiny
\affil[1] {Potsdam Institute for Climate Impact Research, D-14412 Potsdam, Germany}
\affil[2] {Institut f\"ur Theoretische Physik, Justus-Liebig-Universit\"at Giessen, D-35392 Giessen, Germany}
\affil[3] {Department of Physics, Bar-Ilan University, Ramat Gan 52900, Israel}

\begin{document}

\maketitle

\begin{abstract}
The El~Ni\~no Southern Oscillation (ENSO) is the most important driver of climate variability and can trigger extreme weather events and disasters in various parts of the globe. Recently we have developed a network approach, which allows forecasting an El~Ni\~no event about 1 year ahead \cite{Ludescher2013}. 
Here we communicate that since 2012 this network approach, which does not involve any fit parameter, correctly predicted the absence of El~Ni\~no events in 2012, 2013 and 2017 as well as the onset of the large  El~Ni\~no event that started in 2014 and ended in 2016 \cite{Ludescher2014}. Our model also correctly forecasted the onset of the last El Ni\~no event in 2018. 
In September 2019, the model indicated the return of El~Ni\~no in 2020 with an 80\% probability.  
\end{abstract}

\section{The El~Ni\~no Southern Oscillation}
The El~Ni\~no-Southern Oscillation (ENSO) phenomenon \cite{Clarke08, Sarachik10, Power2013, Dijkstra2005, Wang2017,Timmermann2018} 
can be perceived as a self-organized dynamical see-saw pattern in the Pacific ocean-atmosphere system, featured by rather irregular warm (``El~Ni\~no'') and cold (``La Nina'') excursions from the long-term mean state. The ENSO phenomenon  is  quantified by the  Oceanic Ni\~no Index (ONI), which is based on the  average of the sea-surface temperatures (SST) in the Ni\~no3.4 region in the Pacific (see Fig. 1). 
The ONI is
defined as the three-month running-mean SST anomaly in the Ni\~no3.4 region and is a principal measure for monitoring, assessing and predicting ENSO. We will refer to the ONI also as NINO3.4 index.

An El~Ni\~no-episode is said to occur when the index is 0.5°C above the average for at least 5 months. Table 1 shows the ONI between 2012 and present, as communicated by the  National Oceanic and Atmospheric Administration (NOAA) \cite{NOAA}.  
The El~Ni\~no periods are in boldface. The table shows that  there were no El~Ni\~no events in 2012, 2013 and 2017. Between 2011 and present, there were 2 El Ni\~no events. One started in late 2014 and ended in the middle of 2016, the other one started in late 2018 and ended in the middle of 2019. It seems  unlikely that the conditions for an El~Ni\~no episode will be met again in late 2019 \cite{IRI, NOAA}.

\begin{figure}[]
\begin{center}
\includegraphics[width=10cm]{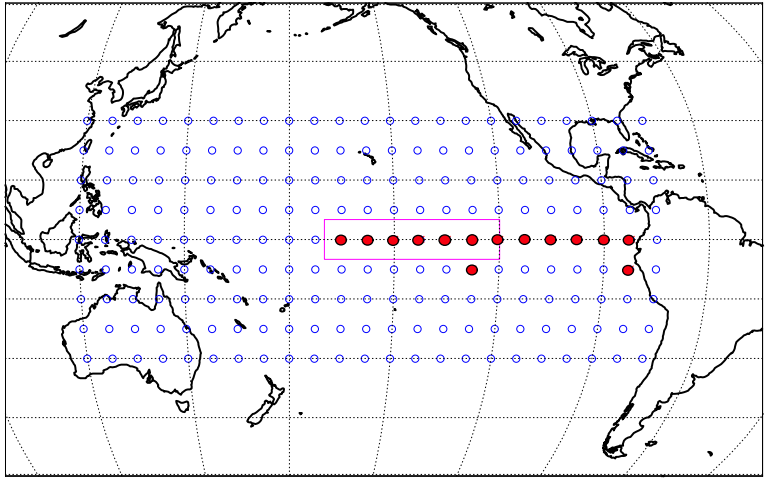}
\caption{The  ONI and the ``climate network''. The network consists of 14 grid
points  in the ``El~Ni\~no basin''  (solid red symbols)  and 193 grid points outside this
domain (open symbols). The red
rectangle denotes the area where the ONI (Ni\~no3.4 index) is measured. The grid points are considered as the nodes of the climate network that we use here to forecast El~Ni\~no events. Each node inside the El~Ni\~no basin is linked to each node outside the basin. The nodes are
characterized by their surface air temperature (SAT), and the link strength between the nodes is determined from their cross-correlation (see below). Figure from \cite{Ludescher2013}.}
\label{fig1}
\end{center}
\end{figure}

\begin{table}[ht]

\renewcommand\arraystretch{1.03}

\centering
\begin{tabular}{ | c | c c c c c c c c c c c c |}
\hline 
\hline 
\rule[2.1mm]{0mm}{2.1mm} 
Year & DJF & JFM & FMA & MAM & AMJ & MJJ & JJA & JAS & ASO & SON & OND & NDJ \\[0.5ex]
\hline

2012 & -0.8 & -0.6 & -0.5 & -0.4 & -0.2 & 0.1 & 0.3 & 0.3 & 0.3 & 0.2 & 0.0 & -0.2 \\
2013 & -0.4 & -0.3 & -0.2 & -0.2 & -0.3 & -0.3 & -0.4 & -0.4 & -0.3 & -0.2 & -0.2 & -0.3 \\
2014 & -0.4 & -0.4 & -0.2 & 0.1 & 0.3 & 0.2 & 0.1 & 0.0 & 0.2 & 0.4 & \bf{0.6} & \bf{0.7} \\
2015 & \bf{0.6} & \bf{0.6} & \bf{0.6} & \bf{0.8} & \bf{1.0} & \bf{1.2} & \bf{1.5} & \bf{1.8} & \bf{2.1} & \bf{2.4} & \bf{2.5} & \bf{2.6} \\
2016 & \bf{2.5} & \bf{2.2} & \bf{1.7} & \bf{1.0} & \bf{0.5} & 0.0 & -0.3 & -0.6 & -0.7 & -0.7 & -0.7 & -0.6 \\
2017 & -0.3 & -0.1 & 0.1 & 0.3 & 0.4 & 0.4 & 0.2 & -0.1 &-0.4 &-0.7 &-0.9 &-1.0 \\
2018 &-0.9  &-0.8  &-0.6&-0.4&-0.1&0.1&0.1&0.2&0.4&\bf{0.7}&\bf{0.9}&\bf{0.8}\\
2019 &\bf{0.8}&\bf{0.8}&\bf{0.8}&\bf{0.8}&\bf{0.6}&\bf{0.5}&0.3&0.1 &&&&\\
\hline

\end{tabular}
\caption{Oceanic El~Ni\~no Index (ONI) 2012 - present. Data from \cite{NOAA}.}
\label{table1}
\end{table}

\section{The forecasting algorithm}
Since  strong El~Ni\~no episodes can wreak havoc in various parts of the world (through 
extreme weather events and other environmental perturbations)  \cite{Wen2002,Corral10,Donnelly07,Kovats03,Davis2001},
early-warning schemes based on robust scientific evidence are highly desirable.
Sophisticated global climate models taking into account the atmosphere-ocean coupling as well as statistical approaches like the dynamical systems schemes approach, autoregressive models and pattern-recognition techniques have  been proposed to forecast the pertinent index with lead times between 1 and 24 months 
\cite{Clarke08,Cane86,Latif94,Tziperman97,Kirtman98,Landsea00,Kirtman03,Fedorov03,Muller04,Chen04,Palmer06,Luo08,Yeh09,Chekroun11,Galanti03,Chen08,Penland1995,Chapman2015, Nootboom2018, Meng2018}. 

Unfortunately, the forecasting methods in use so far have quite limited anticipation power. 
In particular, they generally fail to overcome the so-called ``spring barrier'' (see, e.g., \cite{Webster1995,Goddard2001}), which shortens 
their warning time to around 6 months.

To resolve this problem, we have recently introduced an alternative forecasting approach \cite{Ludescher2013}  based on complex-networks analysis \cite{Kurths,Tsonis2006,Yamasaki2008,Gozolchiani2011,Dijkstra2019} that can considerably shift the probabilistic prediction horizon.  
The approach exploits the remarkable observation that a large-scale cooperative mode linking the ``El~Ni\~no basin'' (i.e., the equatorial Pacific corridor) and the rest of the Pacific ocean (see Fig. 1) builds up in the calendar year before an El~Ni\~no event. An appropriate measure for the emerging cooperativity can be derived from the time evolution of the teleconnections (``links``) between the atmospheric temperatures at the grid points (''nodes``) inside and outside of the El~Ni\~no basin. The strengths of those links are represented by the values of the respective cross correlations (see Data and Methods Section). 
The crucial entity is the mean link strength $S(t)$ as obtained by averaging over all individual links in the network at a given instant $t$ (for details, see \cite{Ludescher2013} and Data and Methods Section). $S(t)$ rises when the cooperative mode builds up and drops again when this mode collapses rather conspicuously with the onset of the El~Ni\~no event. The rise of $S(t)$ in the year before an El~Ni\~no event starts serves as a precursor for the event. 

For the sake of concrete forecasting, we  employed in \cite{Ludescher2013} high-quality  atmospheric temperature data for the 1950-2011 period. The optimized algorithm (see Data and Methods Section) involves an empirical decision threshold $\Theta$.  Whenever $S$ crosses $\Theta$ from below while the most recent  ONI is below 0.5°C, 
the algorithm sounds an alarm and predicts an El~Ni\~no inception in the following year. 
For obtaining and testing the appropriate thresholds,  we divided the data into two halves. In the first part (1950-1980), which represents the learning phase, all thresholds above the temporal mean of $S(t)$ are considered
and the optimal ones, i.e., those that lead to the best predictions in the learning phase, have been determined. 
We found that $\Theta$-values between $2.815$ and $2.834$ lead to the best performance \cite{Ludescher2013}, with a false alarm rate of 1/20. In the second part of the data set (1981-2011), which represents the prediction (hindcasting) phase, the performance of  these thresholds has been tested. We found that  the thresholds between 2.815 and 2.826 gave the best 
results (see Fig. 2, where $\Theta=2.82$). The alarms were correct in 75\% and the non-alarms in 86.4\% of 
the cases. For $\Theta$-values between $2.827$ and $2.834$, the performance was only slightly weaker.

\section{Forecasting the next El~Ni\~no (2011 - present)}
Based on this hindcasting capacity, the approach already has been used in \cite{Ludescher2014}
to extend the prediction phase from the end of 2011 until November 2013.  We like to emphasize that in the forecasting phase, the algorithm does not contain any fit parameter, since the decision thresholds are fixed and  the mean link strengths only depend on the atmospheric temperature data.

In 2011 and 2012 (see Fig. 3),  $S(t)$ did not cross the threshold from below, this way correctly forecasting the absence of El~Ni\~no events in both 2012 and 2013. These predictions, made by the end of 2011 and 2012, respectively, are not trivial.  For example, as late as August 2012, the CPC/IRI Consensus Probabilistic ENSO forecast yielded a 3 in 4 likelihood 
for an El~Ni\~no event in 2012, which turned out to be incorrect only a few months later \cite{IRI,NOAA}. 

In 2013, our algorithm predicted the return of an El~Ni\~no event in 2014, since, in
September 2013,  $S(t)$ transgressed the alarm threshold band while the last available ONI (JJA 2013) was below 0.5°C, indicating the return of El~Ni\~no in 2014 (see Fig. 3).  This early prediction was correct (see Table 1):  The El~Ni\~no event started in November 2014 (and ended in May 2016).

\begin{figure}[]
\begin{center}
\includegraphics[width=14cm]{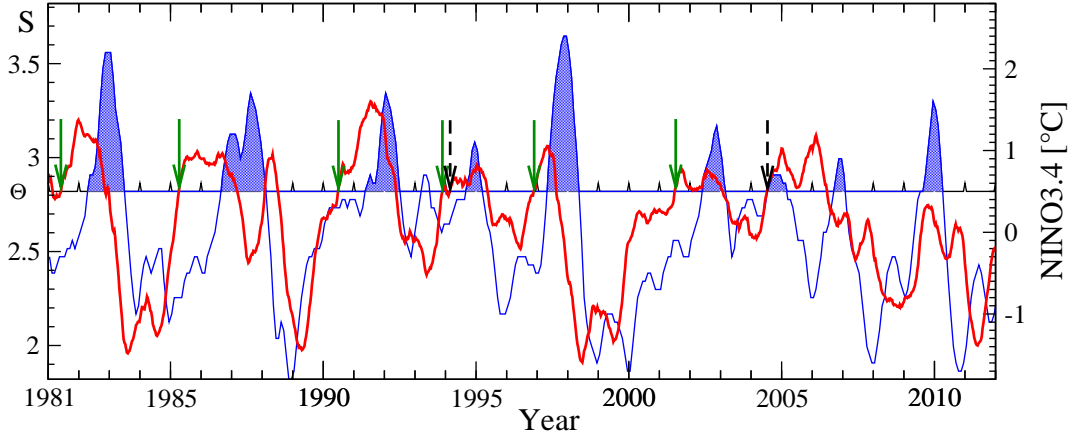}
\caption{ The forecasting scheme.  We compare the average link strength $S(t)$   
in the climate network (red curve) with a decision threshold $\Theta$ (horizontal line, here $\Theta = 2.82$), (left scale), and the 
standard NINO3.4 index (ONI), (right scale), between January  1981 and December 2011. 
When the link strength crosses the threshold from below, and the last available ONI is below 0.5°C,
we give an alarm and predict that an El~Ni\~no episode will start in the following calendar year. 
The El~Ni\~no episodes (when the NINO3.4 index is above $0.5^oC$ for at least 5 mo) are shown by the solid blue areas. 
Correct predictions are marked by green arrows and false alarms by dashed arrows. The alarm in July 2004 must be regarded as a false alarm since the last available ONI (AMJ 2004) was below  0.5°C. 
Between 1981 and 2011, there were 9 El Ni\~no events. The algorithm generated 8 alarms, and 6 were correct. 
In the whole period between 1981 and October 2019, there were 11 El Ni\~no events. The algorithm generated 10 alarms, and  8 were correct.}
\label{fig2}
\end{center}
\end{figure}

\begin{figure}[]
\begin{center}
\includegraphics[width=12cm]{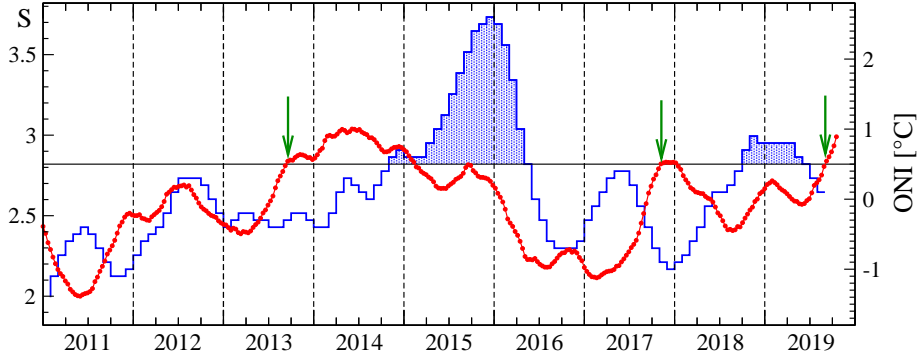}
\caption{ 
The forecasting phase. Same as Fig. 2 but for the 
period between January 2011 and 29 October 2019.  Note that at present 
(October 2019), the ONI curve is only known up to August 2019 (JAS 2019). In September 2019, $S(t)$ transgressed all thresholds. Since also the last ONI, for JJA, is below 0.5°C (0.3°C),
this indicates the return of El Ni\~no in 2020. Since in the past (1981-August 2019) our algorithm generated 10 alarms and 8 were correct, the likelihood of the event is 80 \%.
}
\label{fig3}
\end{center}
\end{figure}

Here, we extend the prediction period further, until 29 October 2019 (present), see Fig. 3.  In  2014, the mean link strength $S(t)$ did not cross any decision thresholds from below, this way correctly forecasting the absence of a new El Ni\~no in 2015. 

In 2015, since the ONI was above 0.5°C for all months, our algorithm did not deliver an alarm. This turned out to be correct since there was no onset of a new El Ni\~no in late 2016.   
 
In 2016 the mean link strength $S(t)$ was well below the decision thresholds. Accordingly, at the end of 2016, our algorithm  predicted the absence of an El~Ni\~no event in 2017. Also this prediction is far from being trivial. As late as April/May 2017, the major forecast schemes (ECMWF,  CPC/IRI Model based ENSO forecast, CPC/IRI Official Probabilistic ENSO forecast) 
predicted an event in 2017 with 3 in 4, 2 in 3, and 1 in 2 likelihoods.

In November 2017,   $S(t)$ transgressed from below the lower threshold band between $S=2.815$ and 2.826. Since the last ONI,  for ASO 2017, was below 0.5°C (-0.4°C),
this indicated the return of El~Ni\~no in 2018 (see Fig. 3), now  with a 7 in 9 likelihood (78\%). The prediction turned out to be correct. Indeed, the forecasted El Ni\~no started in October 2018 and ended in June 2019. 
  
Before coming to the next forecasts,   let us discuss the probability that the same or better outcomes can be obtained by simply guessing. In the 69 years between 1950 and 2018, 23 El Ni\~nos started. Accordingly, the probability that an El Ni\~no starts in a certain year is 1/3. The probability to correctly forecast the El Ni\~nos between 2012 and 2018 is, therefore,
$p=(1/3)^2 (2/3)^5\cong 0.015$. Similarly, one can obtain the probability that in the whole hindcasting and forecasting period, between 1982 and 2018, random guessing would yield better or equal forecasts than our algorithm to be $p\cong 2.2\cdotp10^{-5}$.
  
In 2018, $S(t)$ did not cross any of the decision thresholds from below, this way forecasting at the end of 2018 that in late 2019, with 89\% probability, no new El Ni\~no will start. Right now, this seems likely to be correct. Also, the current official CPC/IRI forecast \cite{IRI} suggests (with 73\% probability) that there will be no  new El Ni\~no  this year.
  
Finally, we find that in September 2019, $S(t)$ transgressed all thresholds. Since the last ONI, for JJA 2019, is below 0.5°C (0.3°C),
this indicates the return of El Ni\~no in 2020. Since in the past (1981-August 2019) our algorithm generated 10 alarms and 8 were correct, now the likelihood of the El~Ni\~no event is 80 percent.
 
We like to note that our algorithm only can  warn of  the El~Ni\~no event next year but not forecast its strength and duration. 
Accordingly, we do not know if the next   El~Ni\~no will be strong or not. We hope that in the near future, reasonable early forecasts of the El Ni\~no magnitudes
will be available by a complexity based approach using information entropy\cite{Meng2019}.
 
An average El~Ni\~no event typically increases the climate anomaly (deviation of global mean surface temperature from pre-industrial level) by about 0.1°C. 
This suggests that a strong El~Ni\~no event in late 2020 can make 2021 a new record year, since air temperature rise lags Pacific warming by about 3 months. 

\section{Data and Methods}

This Section follows closely \cite{Ludescher2014}.
For the prediction of  El~Ni\~no events or non-events, we use the cooperative 
behavior of the atmospheric temperatures in the Pacific as a precursor. To obtain a measure for the 
cooperativity we consider the daily surface atmospheric temperatures (SAT) between June 1948 and October 2019 
temperature data at the grid points (''nodes'') of a Pacific network, see Fig. \ref{fig1}.   

We analyse the time evolution of the teleconnections (``links'') between  the temperatures at nodes $i$ inside the  ``El~Ni\~no basin'' and 
nodes $j$ outside the basin. The strengths of these links are represented by the strengths of the cross correlations between the temperature records at 
these sites \cite{Gozolchiani2011}. 

The prediction algorithm \cite{Ludescher2013, Ludescher2014}  is as follows:

(1) At each node $k$ of the network shown in Fig. 1, the daily
atmospheric temperature anomalies $T_k(t)$ (actual temperature
value minus climatological average  for each calendar day, see below) at
the surface level is determined.  
For the calculation of the climatological average,  the leap days have been removed.
The data have been obtained from the
National Centers for Environmental Prediction/National Center
for Atmospheric Research Reanalysis I project \cite{reanalyis1,reanalyis2}.

(2) For obtaining the time evolution of the strengths of the links
between the nodes $i$ inside the El~Ni\~no basin and the nodes $j$
outside we compute, for each 10th day $t$ in the considered time
span between January 1950 and November 2017, the time-delayed cross-correlation
function defined as
\begin{equation}
 C_{i,j}^{(t)}(-\tau)=\frac{\langle T_i(t)T_j(t-\tau)\rangle-\langle T_i(t)\rangle\langle T_j(t-\tau)\rangle}{\sqrt{\langle(T_i(t)-\langle T_i(t)\rangle)^2 \rangle}\cdot\sqrt{\langle(T_j(t-\tau)-\langle T_j(t-\tau)\rangle)^2 \rangle}}
\end{equation}
and
\begin{equation}
 C_{i,j}^{(t)}(\tau)=\frac{\langle T_i(t-\tau)T_j(t)\rangle-\langle T_i(t-\tau)\rangle\langle T_j(t)\rangle}{\sqrt{\langle(T_i(t-\tau)-\langle T_i(t-\tau)\rangle)^2 \rangle}\cdot\sqrt{\langle(T_j(t)-\langle T_j(t)\rangle)^2 \rangle}}
\end{equation}
where the brackets denote an average over the past 365 d, according to
\begin{equation}
 \langle f(t) \rangle = \frac{1}{365} \sum_{m=0}^{364} f(t-m). 
\end{equation}
We consider time lags $\tau$ between 0 and
200 d, where a reliable estimate of the background noise level can
be guaranteed.

(3) We determine, for each point in time $t$, the maximum, the
mean, and the standard deviation around the mean of the absolute value of the cross-correlation function
 $|C_{ij}^{(t)}(\tau)|$
 and define the link
strength $S_{ij}(t)$ as the difference between the maximum and the
mean value, divided by the standard deviation. Accordingly, $S_{ij}$ describes the
link strength at day t relative to the underlying background noise (signal-to-noise ratio) and
thus quantifies the dynamical teleconnections between nodes i
and j. 

(4) To obtain the desired mean strength $S(t)$ of the dynamical
teleconnections in the climate network we simply average over
all individual link strengths.

(5) Finally, we compare $S(t)$ with a decision threshold $\Theta$. When the link strength $S(t)$ crosses the
threshold from below and  the last available ONI at that time $t$ is below 0.5°C, we give an alarm and predict that an El~Ni\~no episode will start in the following calendar year.
Since the decision threshold has been fixed in the learning phase between 1950 and 1980, the forecasting algorithm does not contain any fit parameter.

We like to add that for the calculation of the climatological average in the learning phase, all data within this time window 
have been taken into account, while in the prediction phase, only data from the past up to the prediction date have been considered.

Note added: After submitting this article to arXiv, the new information entropy based method proposed by Meng et al. \cite{Meng2019} also forecasted the onset of an El~Ni\~no in 2020 with a magnitude of $1.48\pm0.25°C$. The method correctly hindcasted 9 out of 10 events between 1984 and 2018. 

\subsection*{Acknowledgments}
We thank the “East Africa Peru India Climate Capacities - EPICC” project, which is part of the International Climate Initiative (IKI), supported by the German Federal Ministry for the Environment, Nature Conservation and Nuclear Safety (BMU).

\end{document}